\documentclass[12pt,preprint]{aastex}

\def\etal       {et al.}
\def\ie         {{i.e.},\ }
\def\eg         {{e.g.},\ }
\def\onesig     {$1\sigma$}
\def\twosig     {$2\sigma$}
\def\fivesig    {$5\sigma$}
\def\sixsig     {$6\sigma$}

\def\mira    {$o$~Ceti}
\def\sgra    {Sgr~A*}
\def\msun    {M$_\odot$}

\def\arcm{\ifmmode {' }\else $' $\fi}
\def\arcs{\ifmmode {'' }\else $'' $\fi}
\def\arcmper{\ifmmode \rlap.{'} \else $\rlap{.}' $\fi}
\def\arcsper{\ifmmode \rlap.{''} \else $\rlap{.}'' $\fi}
\def\porm   {\ifmmode\pm\else$\pm$\fi}
\def\kms    {\ifmmode{{\rm ~km~s}^{-1}}\else{~km~s$^{-1}$}\fi}
\def\masy   {\ifmmode{{\rm mas~y}^{-1}}\else{mas~y$^{-1}$}\fi}
\def\micron {\ifmmode{\mu{\rm m}}\else{$\mu$m}\fi}
\def\a      {\ifmmode {\rlap.}^{''}\! \else ${\rlap.}^{''}\!$\fi}
\def\p      {\phantom{0}}
\def\q      {\phantom{00}}

\newbox\grsign \setbox\grsign=\hbox{$>$} \newdimen\grdimen \grdimen=\ht\grsign
\newbox\laxbox \newbox\gaxbox
\setbox\gaxbox=\hbox{\raise.5ex\hbox{$>$}\llap
     {\lower.5ex\hbox{$\sim$}}}\ht1=\grdimen\dp1=0pt
\setbox\laxbox=\hbox{\raise.5ex\hbox{$<$}\llap
     {\lower.5ex\hbox{$\sim$}}}\ht2=\grdimen\dp2=0pt
\def\gax{\mathrel{\copy\gaxbox}}
\def\lax{\mathrel{\copy\laxbox}}

\slugcomment{2002 December 10}

\shorttitle{\sgra\ Position}
\shortauthors{Reid \etal      }

\begin{document}


\title{The Position of Sagittarius A*: II. Accurate Positions
and Proper Motions of Stellar SiO Masers near the Galactic Center}


\author{M. J. Reid}
\affil{Harvard--Smithsonian Center for Astrophysics,
    60 Garden Street, Cambridge, MA 02138}
\email{reid@cfa.harvard.edu}

\author{K. M. Menten}
\affil{Max-Planck-Institut f\"ur Radioastronomie,
       Auf dem H\"ugel 69, D-53121 Bonn, Germany}
\email{kmenten@mpifr-bonn.mpg.de}

\author{R. Genzel, T. Ott \& R. Sch\"odel}
\affil{Max-Planck-Institut f\"ur extraterrestrische Physik,
       Postfach 1312, D-85741 Garching, Germany}
\email{genzel@mpe.mpg.de, ott@mpe.mpg.de, rainer@mpe.mpg.de}

\author{A. Eckart}
\affil{I. Physikalisches Institut, Universit\"at zu K\"oln, 
Z\"ulpicher Strasse 77, 50937 K\"oln, Germany}
\email{eckart@ph1.uni-koeln.de}



\begin{abstract}

\citet{MREG97} accurately determined the position of \sgra\ on an infrared
image, by aligning the infrared image
with positions measured for SiO masers, associated with infrared-bright
evolved stars, at radio wavelengths.
We now report greatly improved radio positions and, for the first 
time, proper motions of many stellar SiO masers at the Galactic Center.  
These positions and motions, coupled with better infrared imaging, 
allow a much improved location of \sgra\ on infrared images.  
With current data, infrared stars can be placed in a reference frame tied 
to \sgra, to an accuracy of $\approx10$~mas in position and 
$\approx1$~\masy\ in motion. The position of \sgra\ is, within uncertainties,
consistent with stellar accelerations and the measured orbital focus of 
the star S2.  The star S2 was observed 
within 16~mas ($\approx130$~AU in projection) of Sgr~A* on 2002 May 02.
Finally, we find that the central stellar cluster moves with \sgra\ to 
within $\approx70$~\kms.  

\end{abstract}



\section{Introduction}

Sagittarius A* (\sgra) is a compact radio source projected toward
the center of the Galaxy \citep{BB74}.  The apparent proper motion of \sgra,
measured against extragalactic radio sources, is consistent with it
being at the dynamical center of the Galaxy \citep{R99,BS99}.
A dense cluster of stars surrounds \sgra, and the {\it relative}
proper motions of these stars increases dramatically with decreasing
projected distance from \sgra\ \citep{EG96,G98}.
These stellar motions exceed $1000$\kms\ at a projected distance of
$0.015$~pc from \sgra.  
Using the radio--infrared alignment presented in this paper,
one star, S2, was found to travel on an elliptical orbit with 
Sgr A* at its focus \citep{S02}.  
This star was moving at $\gax5000$~\kms\ at a distance of 124~AU at 
pericenter passage, implying an enclosed mass of 
$\approx 4 \times 10^6$~\msun. 
The very small peculiar motion of $<20$~\kms\ for \sgra\ is in
stark contrast to the high stellar motions and indicates that
\sgra\ must contain at least $10^3$ to $10^4$~\msun\ \citep{R99,CHL02};
this mass is directly associated with the compact radio source
which is $\lax1$~AU in size \citep{BB98,K98}.  
Together, all of these findings provide compelling
evidence that \sgra\ is a super-massive black hole at the
dynamical center of the Galaxy.

While \sgra\ is a very bright radio source, it is a very dim infrared
source.  Locating \sgra\ on infrared images is complicated by the high
projected density of stars at the Galactic Center and by the 
absence of any obvious infrared counterpart for \sgra.
Fortunately, the Galactic Center stellar cluster contains red giant stars 
that are both strong radio sources (from circumstellar SiO
maser emission) and bright infrared sources.  Because these stars
are visible at both radio and infrared wavelengths, one can use their
radio positions, measured with respect to \sgra, to determine accurate 
``plate'' scales and rotations for infrared images and transfer
the position of \sgra\ to these infrared images.  
In \citet{MREG97} (hereafter Paper~I), 
we developed this technique and determined the 
position of \sgra\ on a diffraction-limited 2.2~\micron\ image of the Galactic
Center region made with the MPE SHARP camera on the ESO 3.5-m New Technology
Telescope in Chile.  
We located \sgra\ with a 95\% (\twosig) confidence
of 60~mas and found no infrared emission above a level of 9~mJy (de-reddened)
toward its position.  This upper limit strongly constrains models for 
the emission from the immediate vicinity of a super-massive black hole.  

Locating \sgra\ on infrared images is important for reasons
other than determining its emission.
Recently it has become possible to observe gravitational acceleration directly,
via orbital curvature in the motions of stars
\citep{G00,E02}.  Uncertainty in the position of the dynamical
center, presumably \sgra, limits estimates of the enclosed mass
determined from observed accelerations.
Approximately two-thirds of a complete orbit has now been observed 
for one star, S2 {\citep{S02}, and calculating orbital parameters is
enhanced by precise knowledge of the dynamical center.  
Ultimately, verification that the dynamical center coincides with the position 
of \sgra\ would link the radiative source, \sgra, directly with the 
gravitational source.  

In the five years since Paper I was published, there have been 
significant advances in both the radio astrometry and infrared
imaging of the Galactic Center region, and it was time to update the
infrared positioning of \sgra.
This paper reports VLA and VLBA observations with greatly improved 
positions and, for the first time, proper motions of stars with
circumstellar SiO masers in the Galactic Center region.
In addition, deep (K$_s=18$~mag), large-field images of the Galactic Center region
at infrared wavelengths were obtained with new instrumentation on
one of the 8.2-m VLT telescopes.  Combining the new radio and
infrared observations has yielded a significant improvement in
the location of \sgra\ in the infrared, helped verify that the radio
source \sgra\ is at the focus of a stellar orbit, and placed the
proper motions of stars in the central cluster in a reference frame
tied to \sgra.

\section{Radio Observations}

SiO masers in the extended atmospheres of late type giants and
supergiants are strongly variable over time scales of $\sim1$~y,
and care must be taken to minimize the effects of
maser variability on proper motion measurements.  
For a typical Mira variable, SiO maser emission occurs at a radius of 
about 8~AU, which corresponds to about 1~mas at the 8~kpc distance of the 
Galactic Center \citep{R93}.  Since individual 
maser features generally persist over most
of a stellar period, but not among different cycles, there are
two approaches to dealing with variability.
One can observe with high angular resolution, \eg using VLBI techniques,
and make measurements only within one cycle, by tracking
{\it individual} maser features within the masing shell.
Alternatively, one can observe with lower angular resolution,
\eg using connected-element interferometers, and make measurements 
over a time-span much longer than the stellar cycle.  
With this approach one does not attempt
to track individual maser features and accepts an intrinsic stellar 
position uncertainty of about 1~mas owing to variations across
the maser shell.  We employed both techniques.
 
Over the period 1995 to 2000 we searched for and mapped
SiO masers associated with late-type stars that are projected near \sgra.  
We used the NRAO\footnote{NRAO is a facility of the National Science Foundation
operated under cooperative agreement by Associated Universities, Inc.} 
VLBA and VLA to measure accurately the relative positions of SiO maser 
stars and \sgra.  
Typically, we obtained positions, accurate to $\approx1$~mas, and proper motions, 
accurate to $\approx1$~mas~y$^{-1}$, for a number of maser stars.  

\subsection{VLBA Observations}

The VLBA observations were conducted on four epochs, spaced approximately
at monthly intervals, on 1996 April 25, May 12, June 15, and July 13.
Strong interstellar scattering toward \sgra, our phase-reference
source, broadens the apparent image of \sgra\ to about 0.7~mas at 43~GHz 
\citep{K98}.  This limits interferometer baselines that can be effectively 
used at this observing frequency to $\lax1500$~km length.
We chose to use only the inner 5 antennas of the VLBA (PT, LA, KP, FD, OV),
which form interferometer baselines shorter than this length
over the entire track on \sgra.

We observed with four 8-MHz bands centered on velocities
of $-20, -65, -120,$ and $-346$~\kms\ with respect to the
local standard of rest (LSR), assuming a rest-line frequency
of 43122.08~MHz for the $v=1, J=1\rightarrow0$ transition of SiO.  
The VLBA correlator generated 128 spectral channels, 
spaced by 62.5~kHz or 0.435~\kms.

The primary beam pattern of a single VLBA antenna at 43~GHz has
a half-width at half-maximum of $\approx 35\arcs$.  Thus,
SiO maser stars within about 15\arcs\ of \sgra\ can be observed
while pointing at \sgra\ with little degradation.  
(We were able to detect strong SiO maser stars out to 42\arcs\ 
with some loss of sensitivity.)
We pointed the VLBA antennas at \sgra\ and cross-correlated the
data only at the position of \sgra.   We later shifted the phase
center of the interferometer data, using special software, 
to positions where SiO masers were known from our earlier 
VLA observations.  In order to allow for this, the correlator 
integration time was limited to 0.4~s, yielding a field-of-view  
that avoids strong ``fringe-rate smearing'' with a radius of 
$\approx12\arcs$.

Calibration was performed in the standard manner outlined in
the NRAO Astronomical Image Processing System (AIPS) 
documentation for spectral-line VLBA observations.
Amplitude calibration employed on-line measurements of system
temperature and standard gain-curves provided by NRAO.
Initial delay and electronic phase calibration was accomplished 
with observations of NRAO~530 at approximately 2-hour intervals.
This allowed the four 8-MHz bands to be combined for the 
\sgra\ observations and interferometer phases to be measured 
every 10 seconds.  Rapid interferometer phase changes are caused 
predominantly by fluctuations in the water vapor content of the 
atmosphere along the lines of sight of the antennas.  By measuring them
on \sgra, we could remove their effects from the data.

Calibrated interferometric data for \sgra\ observations were
then copied and the interferometer phase-center was re-calculated
to allow for the angular offset of an SiO maser star from \sgra.
This could not be accomplished accurately enough with standard 
AIPS tasks (eg, CLCOR), and we used a special AIPS task (PHREF) 
written by A. Beasley.  This task calculated the difference 
of the full interferometer delay and phase 
for the \sgra\ and the SiO maser position.  
The data were corrected for these delay and phase
differences, effectively yielding a calibrated copy of the data as 
if it were correlated at the SiO maser position.   Such data sets were
generated for the SiO masers known at the time of observation
that were within about 12\arcs\ of \sgra: 
IRS~9, IRS~7, IRS~12N, IRS~10EE, and IRS~15NE.  
We also generated such a data set
for \sgra\ as a software control.

Spectral-line image cubes were then made using the AIPS task IMAGR
with maps of $512\times512$ pixels of size of 0.2~mas, 
yielding a field-of-view 
of about 100~mas.  This was more than sufficient to allow for 
uncertainties in the maser positions, which came from the 1995 VLA
observations and were typically 20~mas.  Noise levels in individual 
channel maps were $\approx15$~mJy.  We searched these map cubes 
for SiO maser emission and were able to detect three stars: 
IRS~12N, IRS~10EE, and IRS~15NE.   

Fig.~1 shows spectra and maps of these sources for the first and 
last epoch of our VLBA observations.  
The spectra were constructed by summing the emission over a
region on the sky, typically about 3~mas, that enclosed all of
the maser emission.
The maps were obtained by assigning the maximum brightness 
over the spectral range of the maser emission for each pixel.
This method better displays multiple spectral components with
different strengths than simply summing the emission.
One can clearly see motions of the three stars over the 79~days
between the observations.    Also, one can see that the
SiO emission peaks from IRS~10EE subtend an angle of about 1.8~mas 
on the sky.
Were this star to be at a distance from the Sun of \mira\ ($\approx110$~pc), 
instead of near 8~kpc, the SiO emission would subtend an angle of about
130~mas.  This is a bit larger than, but comparable to, the observed
diameter of the SiO emission from \mira\ of about 90~mas \citep{RM02}.  
Thus, IRS~10EE, and IRS~12N and  IRS~15NE which have more
compact SiO emission, appear to be Mira-like variable stars.

\begin{figure}
\epsscale{1.0}
\plotone{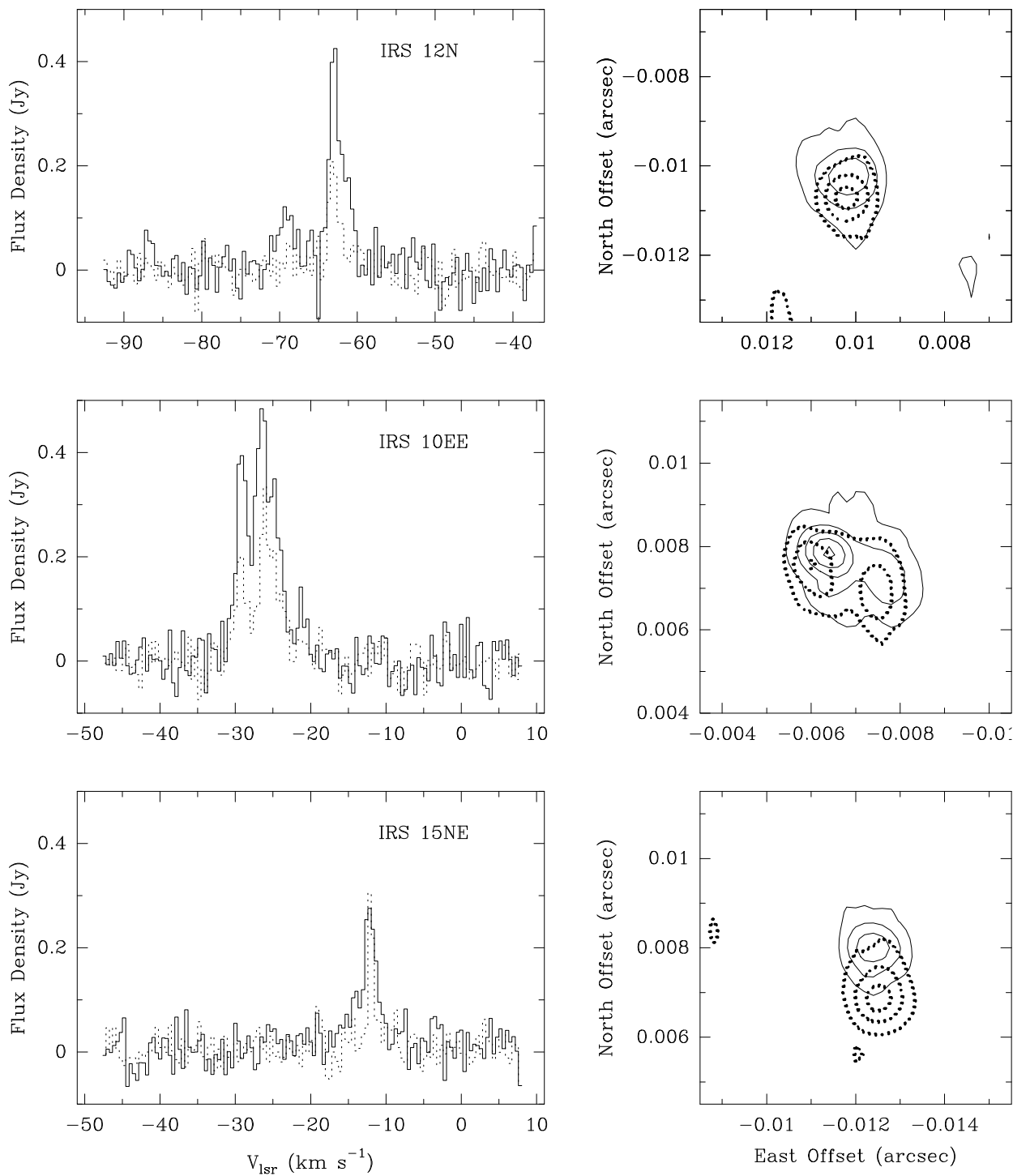}
\caption{Spectra (left panels) and maps (right panels) of three 
SiO masers in the Galactic Center region made with the VLBA.  
Solid and dotted lines are for 
observations on 1996 April 25 and July 13, respectively.  
Map contour levels are multiples of 0.05~Jy~beam$^{-1}$.
The small ($\sim10$~mas) position offsets reflect the \'a priori
position uncertainties before the observations.
Proper motions are clearly visible between the two epochs separated by 97~days.
            \label{fig1}
        }
\end{figure}

While one could obtain a crude estimate of the proper motions of
the SiO maser stars by analyzing Fig.~1, we did the following
to obtain optimum estimates of the position and motion for each star:
\begin{itemize}
\item[] 1) We fit a 2-dimensional Gaussian brightness distribution to each
           spectral channel with detectable SiO emission. 
\item[] 2) On a channel by channel basis, we fit a straight line to the
           east-west and north-south positions as a function of time.
\item[] 3) The channel motions were then averaged, using both variance 
           and uniform weighting.  We adopted the mean of the two methods.
\item[] 4) Proper motion uncertainties were
      estimated by taking larger of the two formal errors (\ie from the
      uniform-weighted solutions), and these were increased by $\sqrt{2}$
      in order to account roughly for the effects of blending of 
      maser spots.
\item[] 5) The formal uncertainties for the mean positions were quite small,
      since we averaged positions of many spectral channels.
      However, the measured SiO shell size appears to be up to 1~mas in 
      radius, and we adopt a position uncertainty of
      0.5~mas at the reference epoch.
\end{itemize}
The proper motion solutions for the three stars detected with the VLBA 
are listed in Table~1.

\begin{deluxetable}{lcrrrrl}
\tabletypesize{\scriptsize}
\tablecaption   {VLBA SiO Maser Proper Motions}
\tablehead{ \colhead{Star} &\colhead{$v_{LSR}$} &\colhead{$\Delta\Theta_x$} 
                           &\colhead{$\Delta\Theta_y$} &\colhead{$\mu_x$}   
                           &\colhead{$\mu_y$} &\colhead{Epoch}     \\
            \colhead{}     &\colhead{(\kms)}    &\colhead{(arcsec)} 
                           &\colhead{(arcsec)}  &\colhead{(mas y$^{-1}$)} 
                           &\colhead{(mas y$^{-1}$)} &\colhead{(y)}       
          }
\tablecolumns{7}
\startdata 
\\
IRS~12N\ ....      &$\p-63$ &$-3.2519\pm0.0005$    &$-6.8814\pm0.0005$ 	&$-0.49\pm0.54$ &$-1.93\pm0.81$ &1996.41   \\
IRS~10EE\ ...     &$\p-27$ &$+7.6837\pm0.0005$    &$+4.2194\pm0.0005$ 	&$+1.14\pm0.38$ &$-1.51\pm0.57$ &1996.41   \\
IRS~15NE\ ...     &$\p-12$ &$+1.2302\pm0.0005$    &$+11.3315\pm0.0005$  &$-0.79\pm0.71$ &$-4.66\pm1.06$ &1996.41   \\
\\
\tablecomments{VLBI position uncertainties reflect the
$\approx0.5$~mas SiO maser shell radius.  
$\Delta\Theta_x$ and $\Delta\Theta_y$ are angular offsets,
and $\mu_x$ and $\mu_y$ are proper motions, relative to  
\sgra\ toward the east and north, respectively, in the J2000 
coordinate system.
}
\enddata
\end{deluxetable}

\subsection{VLA Observations}

The VLA observations were conducted in the A-configuration in 1998 May 
and 2000 October/November.  
Near the Galactic Center, SiO masers are likely detectable 
over a very wide range of velocities, probably exceeding 700~\kms.
However, wide-band observations at the VLA are currently
limited by the correlator and, in order to obtain adequate spectral
resolution, we chose to observe in 6.25~MHz bandwidths (each covering
42~\kms).   With 64 spectral channels per band, this resulted in
a spectral resolution of about 98~kHz or 0.67~\kms.

In 1998, the VLA had 12 antennas operating with 43~GHz receivers.
These antennas were sited so as to cover, somewhat sparsely, the
entire range of A-configuration spacings, and the synthesized
full-width at half-power beamwidth toward \sgra\ was about 
$70 \times 30$~mas elongated in the north-south direction.
We observed with two different equipment setups on different days.
On May 20 and 25, we covered a wide velocity range 
($-365 \rightarrow +365$~\kms), searching for new SiO maser stars.
We used twelve frequency settings, each with two 6.25~MHz bands, 
and observed in one polarization, for a total of about 25 minutes
per setting.  Typical noise levels in a single
spectral channel were 15~mJy.
On May 27 and 31, we concentrated on four 6.25~MHz bands centered at 
$-346, -121, -65, {\rm and} -20$~\kms, which included five SiO maser
stars known at that time.
By restricting the velocity range, we were able to observe longer and in a
dual-circularly polarized configuration in order to increase sensitivity 
and, thus, positional accuracy for these stars.   Although the
weather conditions were somewhat worse on these days, typical noise levels
in a single spectral channel were about 10~mJy.

In 2000, observations were conducted on Oct. 28 \& 30 and Nov. 13 \& 16.
The VLA had all 27 antennas operating with 43~GHz receivers, and
the synthesized beam toward \sgra\ was about $80 \times 40$~mas
elongated in the north-south direction.  
We observed by cycling among seven 6.25 MHz bands centered 
at LSR velocities of $-346, -111, -73, -39, -1, +40$ and $+75$\kms;
the latter six bands covered the LSR velocity range 
$-131 \rightarrow +95$\kms\ with only two small gaps.
This setup allowed the deepest integrations possible for the nine
known, or suspected, SiO maser stars at that time.
We observed in both right and left circular polarization for each band
and obtained 64 spectral channels per band as for the 1998 observations.
The synthesized maps typically had single spectral channel noise levels
of about 5~mJy. 

Initial calibration of all VLA data was done in a standard manner recommended
by the AIPS documentation.
The flux density scale was based on observations of 3C~286, 
assumed to be 1.49~Jy for interferometer baselines shorter than
300~k$\lambda$.  Bandpass calibration was accomplished
with observations of NRAO~530, which had a flux density of 3.6~Jy 
during the 1998 observations and 3.3~Jy during the 2000 observations.
The data were then self-calibrated (amplitude and phase) on \sgra.

When searching for masers with the 1998 data, we made very large images, 
covering about 80\arcs, or most of the primary beam of an individual 
VLA antenna at 43 GHz.
Three new SiO maser stars were discovered within 15\arcs\ of \sgra.  
We searched for masers with the 2000 data within a more restricted 
40\arcs\ region centered on \sgra.  
Typical rms noise levels in these images were 5~mJy,
allowing \sixsig\ detections of 30~mJy.  One new SiO maser star, offset
from \sgra\ by $(+10\arcsper5,-5\arcsper8)$ (E,N),  was discovered.
Including the three SiO masers in Paper~I, these discoveries brought 
the number of detected SiO maser stars within 15\arcs\ of \sgra\ to seven.

Once the approximate location of a maser was known, 
either from previous observations or from the large images, 
we mapped each band with up to four small sub-images centered 
on the known masers with emission in that band.  
We always included a sub-image for \sgra\ at the phase 
center of the interferometric data.  By simultaneously imaging the
stellar SiO masers and the continuum emission from \sgra, 
the strong continuum emission from \sgra\ did not degrade the
detections of relatively weak SiO masers far from \sgra.

We obtained a single position for each star at each observing epoch in the
following manner:
\begin{itemize}
\item[] 1) We fit a 2-dimensional Gaussian brightness distribution to each
           spectral channel with detectable SiO emission. 
\item[] 2) The fitted positions from each observing season 
       were then averaged, using variance weighting, 
      to obtain a best stellar position and estimated uncertainty.
\item[] 3) Since the correlator model for the VLA neglects terms
      for annual aberration \citep{F92}, we calculated the effect this
      had on the {\it relative} position of a star and \sgra, and corrected
      the stellar positions.  For stars within about 15~\arcs\ of \sgra, these
      corrections were less $\lax1$~mas.
\end{itemize}

We list the positions of the SiO maser stars, relative to \sgra, in
Table~2.  In addition to the 1998 and 2000 VLA observations reported here, 
this table includes the 1995 VLA observations reported in Paper I 
and a single position for the 1996 VLBA observations (see \S 2.1) 
determined for an epoch near the center of those observations.  
Since stellar SiO masers are variable in
strength over the period of the stellar pulsation ($\sim1$~y) 
and the sensitivity of each epoch's data differed somewhat, only the
strongest masers were detected at all epochs.

\begin{deluxetable}{lrrrll}
\tabletypesize{\scriptsize}
\tablecaption   {SiO Maser Astrometry}
\tablehead{ \colhead{Star} &\colhead{$v_{LSR}$} &\colhead{$\Delta\Theta_x$} &\colhead{$\Delta\Theta_y$} 
                           &\colhead{Epoch}     &\colhead{Telescope}   \\
            \colhead{}     &\colhead{(\kms)}    &\colhead{(arcsec)}         &\colhead{(arcsec)}      
                           &\colhead{(y)}       &\colhead{}
          }
\tablecolumns{6}
\startdata 
\\
IRS~9 \ .....     &$-340$  &$5.6515\pm0.0048$	 &$-6.3589\pm0.0060$    &1998.39   &VLA\\
		  &	   &$5.6501\pm0.0007$    &$-6.3509\pm0.0013$    &1998.41   &VLA\\
		  &	   &$5.6589\pm0.0011$    &$-6.3454\pm0.0017$    &2000.85   &VLA\\
\\
IRS~7 \ .....     &$-120$  &$0.0403\pm0.0080$    &$5.5829\pm0.0130$     &1995.49   &VLA\\   
                  &        &$0.0387\pm0.0023$    &$5.5676\pm0.0049$     &1998.39   &VLA\\
                  &        &$0.0378\pm0.0043$    &$5.5495\pm0.0014$     &1998.41   &VLA\\
                  &        &$0.0342\pm0.0016$    &$5.5414\pm0.0030$     &2000.85   &VLA\\
\\
IRS~12N\ ....      &$\p-63$ &$-3.2519\pm0.0005$    &$-6.8814\pm0.0005$ 	&1996.41   &VLBA\\
                  &        &$-3.2536\pm0.0005$    &$-6.8876\pm0.0006$ 	&1998.39   &VLA\\
                  &        &$-3.2543\pm0.0010$    &$-6.8860\pm0.0011$ 	&1998.41   &VLA\\
                  &        &$-3.2554\pm0.0009$    &$-6.8936\pm0.0012$ 	&2000.85   &VLA\\
\\
IRS~28\ .....     &$\p-55$ &$10.4702\pm0.0030$   &$-5.7884\pm0.0050$ 	&1998.41   &VLA\\
                  &        &$10.4693\pm0.0010$   &$-5.7956\pm0.0024$ 	&2000.85   &VLA\\
\\
IRS~10EE\ ...     &$\p-27$ &$7.6821\pm0.0030$    &$4.2125\pm0.0050$ 	&1995.49   &VLA\\
                  &        &$7.6837\pm0.0005$    &$4.2194\pm0.0005$ 	&1996.41   &VLBA\\
                  &        &$7.6841\pm0.0005$    &$4.2146\pm0.0009$ 	&1998.39   &VLA\\
                  &        &$7.6837\pm0.0005$    &$4.2157\pm0.0005$ 	&1998.41   &VLA\\
                  &        &$7.6845\pm0.0005$    &$4.2099\pm0.0005$ 	&2000.85   &VLA\\
\\
IRS~15NE\ ...     &$\p-12$ &$1.2256\pm0.0140$    &$11.3108\pm0.0210$    &1995.49   &VLA\\
                  &        &$1.2302\pm0.0005$    &$11.3315\pm0.0005$    &1996.41   &VLBA\\
                  &        &$1.2249\pm0.0017$    &$11.3193\pm0.0019$    &1998.39   &VLA\\
                  &        &$1.2270\pm0.0005$    &$11.3201\pm0.0006$    &1998.41   &VLA\\
                  &        &$1.2228\pm0.0011$    &$11.3024\pm0.0025$    &2000.85   &VLA\\
\\
SiO--6\ .....     &$\p+52$ &$35.1982\pm0.0090$   &$30.6567\pm0.0140$    &1995.49   &VLA\\
                  &        &$35.2207\pm0.0029$   &$30.6537\pm0.0050$    &1998.39   &VLA\\
                  &        &$35.2323\pm0.0006$   &$30.6593\pm0.0010$    &2000.85   &VLA\\
\\
SiO--11\ ....     &$\p+72$ &$1.7121\pm0.0040$    &$40.2614\pm0.0060$    &1995.49   &VLA\\
                  &        &$1.7379\pm0.0023$    &$40.2681\pm0.0032$    &1998.39   &VLA\\
                  &        &$1.7401\pm0.0005$    &$40.2794\pm0.0011$    &2000.85   &VLA\\
\\
IRS~17\ .....     &$\p+75$ &$+13.1501\pm0.0026$   &$+5.5651\pm0.0025$   &2000.85   &VLA\\
\\
SiO--12\ ....     &$\p+82$ &$-18.8645\pm0.0190$   &$42.4895\pm0.0290$   &1995.49   &VLA\\
                  &        &$-18.8235\pm0.0028$   &$42.4686\pm0.0032$   &2000.85   &VLA\\
\\
\tablecomments{VLBA positions are reported at a single reference epoch.
VLA data have been corrected for missing annual aberration terms.
$\Delta\Theta_x$ and $\Delta\Theta_y$ are angular offsets,
and $\mu_x$ and $\mu_y$ are proper motions, relative to  
\sgra\ toward the east and north, respectively, in the J2000 
coordinate system.
}
\enddata
\end{deluxetable}

We constructed spectra at the pixel of peak brightness for SiO masers
detected in the 1998 and 2000 VLA observations.  These spectra are
displayed in Fig.~2.    Most of these SiO spectra are close to
those expected from Mira variables located at the distance 
of the the Galactic Center.  They show emission covering a velocity
range of $5\rightarrow10$~\kms\ and strong variability over
timescales of years.  

The most unusual SiO maser source in the Galactic Center region is IRS~7.  
It has a supergiant luminosity and SiO maser features spread over
$\approx20$~\kms.  The peak strength of the maser emission from IRS~7
seems to have decreased since our first detection at 1995.49
of 0.4~Jy (see Paper~I) to 0.1~Jy at 1998.40 and 0.05~Jy at 2000.85.
Also, the velocity of the {\it peak} maser feature changed significantly
from about $-124$~\kms\ at 1995.49 and 1998.40 to about $-103$\kms\ 
at 2000.85.  If IRS~7 is similar to the nearby supergiant star
VY~CMa, it may have an SiO maser shell that is much 
larger than typical Miras.  VY~CMa has SiO masers
spread over about 100~mas, with the strongest features contained
within a region of about 50~mas \citep {RM03}.  
Since VY~CMa is at a distance of
about 1.5~kpc \citep{LR78}, were it at the distance of the Galactic Center 
(8~kpc), the strongest maser features would span about 10~mas on the sky.
Thus, the positions determined from a single SiO maser feature in IRS~7 
may not indicate the stellar position to better than about 5~mas.

\begin{figure}
\epsscale{0.95}
\plotone{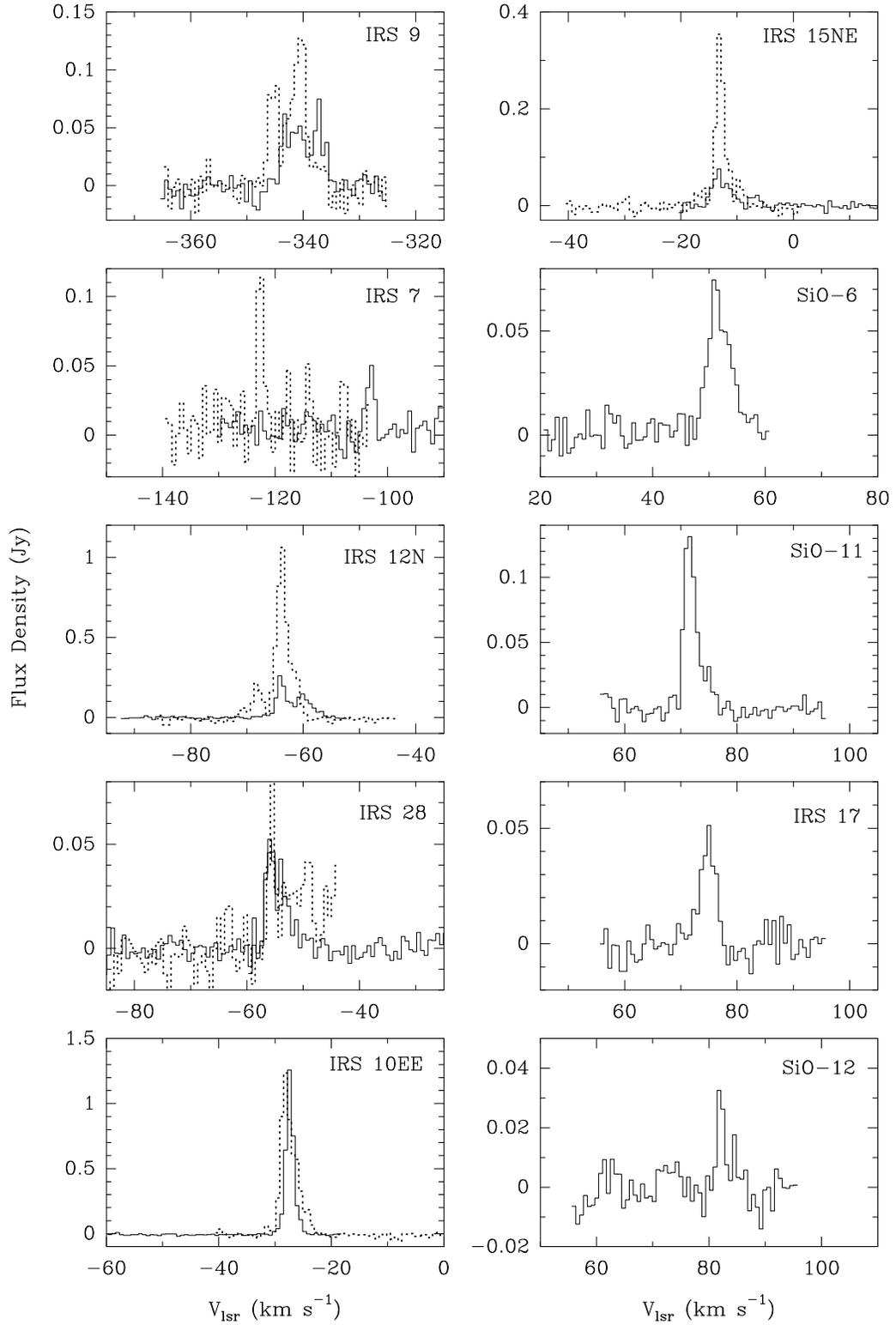}
\caption{Spectra of the SiO masers detected with the VLA during
the 1998 ({\it dotted lines}) and 2000 ({\it solid lines}) observations.  
The year 2000 spectrum of IRS~28 is a 
composite of spectra taken in two overlapping bands.
Strong variations, typical of Mira variables, are seen between the
two epochs 2.4~y apart.
            \label{fig2}
        }
\end{figure}

\subsection{Combined VLA \& VLBA SiO Maser Proper Motions}

We determined stellar proper motions by fitting a straight line to
the positions as a function of time from all of the available data 
compiled in Table~2.  
These proper motion fits are given in Table~3, and those from more than 
two epochs are displayed graphically in Fig.~3.  
The reference epoch for the proper motion 
solution was chosen as the variance-weighted mean epoch for each star,
in order to obtain uncorrelated position and motion estimates.
However, since the estimated uncertainties for individual east-west 
and north-south positions were neither identical, nor exactly linearly 
related, we chose a single, average reference epoch for each star 
(instead of a separate reference epoch for the east-west and 
north-south directions), which resulted in slight parameter correlations.

It is interesting to compare the three stars with proper motion estimates from 
both the VLBA observations (spanning 79~d) and the combined VLA/VLBA 
positions (spanning 5 years).  These motion estimates
are essentially independent, since they use different telescopes arrays,
angular resolutions, interferometer correlators, and correlator models.  
The inclusion of the single VLBA position with the VLA positions
does not correlate the two {\it motion} estimates and provides a valuable cross-check
on the relative position measurements of the two telescope arrays.
Note that there is no indication of a relative position offset between the
VLA and the VLBA measurements.
The east-west and north-south motions (in Tables 1 \& 3) for IRS~12N, IRS~10EE,
and IRS~15NE differ typically by about their joint uncertainties.  
Only one component motion difference is greater than \twosig: $0.94\pm0.41$~\masy\ 
for the east-west motion difference of IRS~10EE, which is not unreasonably large.

Assuming a distance of 8.0~kpc to the Galactic Center, the proper motions
can be converted from angular to linear velocities.  Since the line-of-sight
velocity component is also known to within about 5~\kms\ for each star 
from its SiO maser (LSR) velocity, we can also calculate the full 
3-dimensional speed, $V_{total}$, of these stars, which are given in Table~4.  
The estimated 3-dimensional speeds assume that \sgra\ has a zero velocity 
with respect to the LSR.  Since no spectral lines have been detected from \sgra,
there is no {\it direct} evidence supporting this assumption.
However, strong indirect evidence comes from upper limits to the intrinsic 
proper motion of \sgra\ itself, with respect to extragalactic sources 
\citep{R99,BS99}.  After removing the best estimates of the effects of
the Sun's orbit (\ie the motion of the LSR and the peculiar motion of the Sun) 
from the apparent proper motion of \sgra, \sgra\ appears stationary at the 
Galactic Center to within about 20~\kms\ \citep{R99}.  
Thus, \sgra\ seems to be at rest at the dynamical
center of the Galaxy, and it is highly unlikely that \sgra\
has a significant non-zero LSR velocity.
Thus, the 3-dimensional speeds of the nine SiO maser stars in Table~4 are 
almost surely properly referenced to \sgra.  

\begin{figure}
\epsscale{1.0}
\plotone{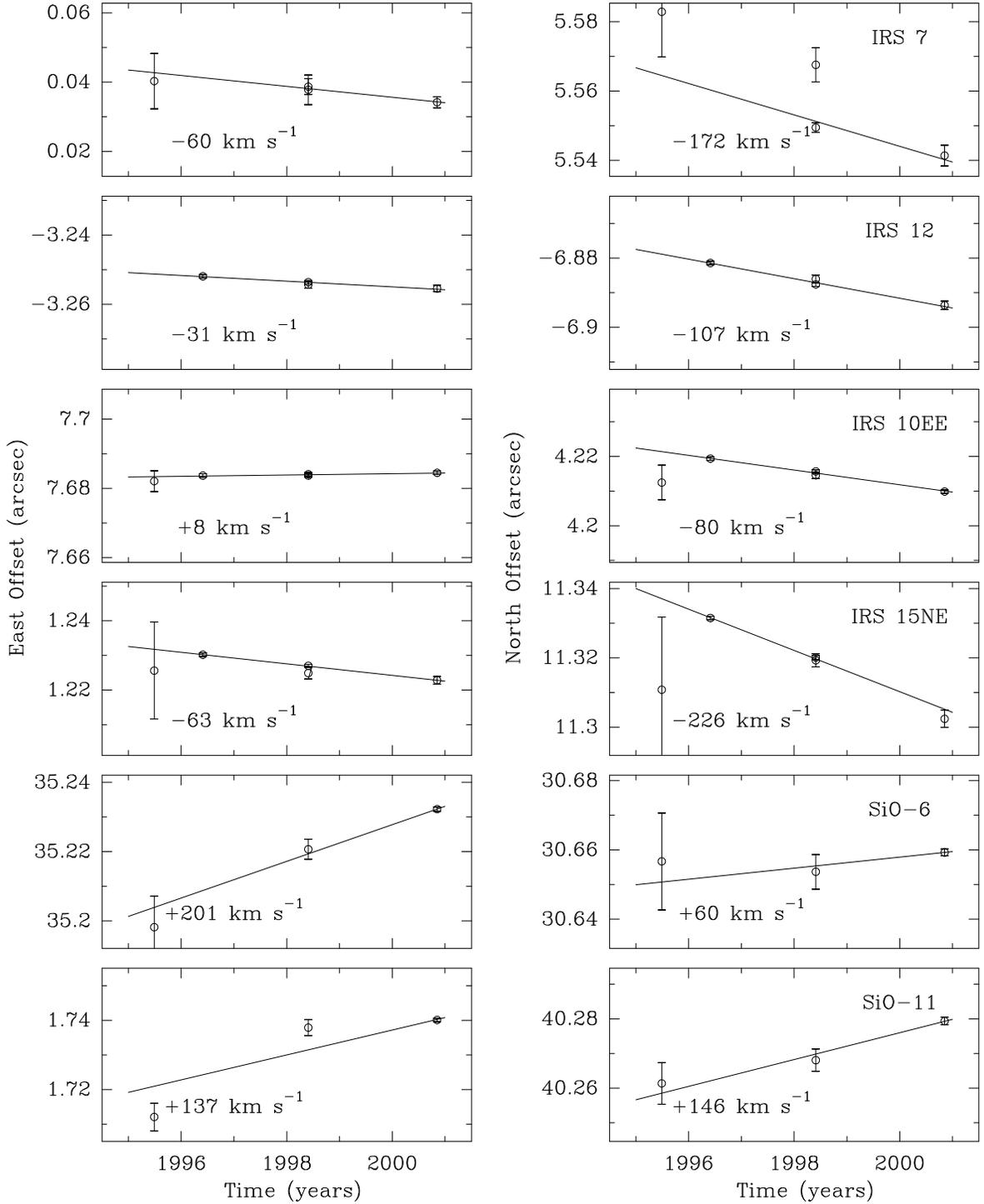}
\caption{Eastward (left panels) and northward (right panels) position
offsets from \sgra\ versus time for the six SiO maser stars 
with detections at three or more epochs.  Solid lines are variance-weighted
best-fit proper motions. The linear speed is indicated in each frame,
assuming a distance of 8.0~kpc.  Star names are indicated in the upper
right corner of the right panels.
            \label{fig3}
        }
\end{figure}

\begin{deluxetable}{lrrrrrll}
\tabletypesize{\scriptsize}
\tablecaption   {Combined VLA \& VLBA Proper Motions}
\tablehead{ \colhead{Star} &\colhead{$v_{LSR}$} &\colhead{$\Delta\Theta_x$} &\colhead{$\Delta\Theta_y$}
                           &\colhead{$\mu_x$}   &\colhead{$\mu_y$}
                           &\colhead{Epoch}     &\colhead{Number} \\
            \colhead{}     &\colhead{(\kms)}    &\colhead{(arcsec)}              &\colhead{(arcsec)}      
                           &\colhead{(mas y$^{-1}$)} &\colhead{(mas y$^{-1}$)}&\colhead{(y)} &\colhead{Epochs}   
          }
\tablecolumns{8}
\startdata 
\\
IRS~9             &$-340$  &$+5.6531\pm0.0006$    &$-6.3493\pm0.0013$   &$+3.60\pm0.53$ &$+2.40\pm1.13$ &1999.24   &2 \\
IRS~7 \ .....     &$-120$  &$+0.0364\pm0.0013$    &$+5.5461\pm0.0043$   &$-1.57\pm0.92$ &$-4.54\pm3.47$ &1999.55   &3 \\
IRS~12N\ ....      &$\p-63$ &$-3.2531\pm0.0003$    &$-6.8853\pm0.0003$ 	&$-0.82\pm0.22$ &$-2.81\pm0.26$ &1997.77   &3 \\
IRS~28\ .....     &$\p-55$ &$+10.4694\pm0.0010$   &$-5.7944\pm0.0022$ 	&$-0.37\pm1.30$ &$-2.95\pm2.27$ &2000.44   &2 \\
IRS~10EE\ ...     &$\p-27$ &$+7.6840\pm0.0003$    &$+4.2150\pm0.0003$ 	&$+0.20\pm0.16$ &$-2.12\pm0.20$ &1998.52   &4 \\
IRS~15NE\ ...     &$\p-12$ &$+1.2283\pm0.0003$    &$+11.3249\pm0.0004$  &$-1.66\pm0.24$ &$-5.96\pm0.35$ &1997.54   &4 \\
SiO--6\ .....     &$\p+52$ &$+35.2317\pm0.0006$   &$+30.6591\pm0.0010$  &$+5.30\pm0.99$ &$+1.59\pm1.64$ &2000.73   &3 \\
SiO--11\ ....     &$\p+72$ &$+1.7390\pm0.0018$    &$+40.2780\pm0.0010$  &$+3.60\pm2.15$ &$+3.86\pm0.91$ &2000.51   &3 \\
IRS~17\ .....     &$\p+75$ &$+13.1501\pm0.0026$   &$+5.5651\pm0.0025$   &...            &...            &2000.85   &1 \\
SiO--12\ ....     &$\p+82$ &$-18.8241\pm0.0028$   &$+42.4689\pm0.0032$  &$+7.65\pm3.58$ &$-3.90\pm5.44$ &2000.77   &2 \\
\\
\tablecomments{  
For sources with VLBA detections, only a single position was used when fitting
for proper motions.  VLA data have been corrected for missing annual aberration terms in the correlator model \citep{F92}.
$\Delta\Theta_x$ and $\Delta\Theta_y$ are angular offsets,
and $\mu_x$ and $\mu_y$ are proper motions, relative to  
\sgra\ toward the east and north, respectively, in the J2000 
coordinate system.
}
\enddata
\end{deluxetable}

\begin{deluxetable}{lrrrrrrrrr}
\tabletypesize{\scriptsize}
\tablecaption   {3-Dimensional Stellar Motions \& Enclosed Mass Limits}
\tablehead{ \colhead{Star} &\colhead{$v_{LSR}$} 
           &\colhead{$v_x$}     &\colhead{$v_y$}     
           &&\colhead{$V_{total}$}     &\colhead{$V_{min}$}
	   &&\colhead{R$_{proj}$}          &\colhead{M$_{encl}$}     \\
            \colhead{}     &\colhead{(\kms)}    
           &\colhead{(\kms)}    &\colhead{(\kms)}   
           &&\colhead{(\kms)}          &\colhead{(\kms)}
           &&\colhead{(pc)}            &\colhead{(\msun)}
           }
\tablecolumns{10}
\startdata 
\\
IRS~9\ ......     &$-340\pm5$   &$+137\pm\p20$ &$+91\pm\p43$    &&$378$ &$344$  &&0.33 &$>4.5\times10^6$ \\
IRS~7 \ .....     &$-120\pm5$   &$-60\pm\p35$  &$-172\pm132$    &&$218$ &$110$  &&0.22 &$>3.0\times10^5$ \\
IRS~12N\ ....     &$\p-63\pm5$  &$-31\pm\q8$   &$-107\pm\p10$   &&$128$ &$103$  &&0.30 &$>3.6\times10^5$ \\
IRS~28\ .....     &$\p-55\pm5$  &$-14\pm\p49$  &$-112\pm\p86$   &&$125$ &$45$   &&0.47 &$>1.1\times10^5$ \\
IRS~10EE\ ...     &$\p-27\pm5$  &$+8\pm\q6$    &$-80\pm\q8$     &&$85$  &$67$   &&0.34 &$>1.8\times10^5$ \\
IRS~15NE\ ...     &$\p-12\pm5$  &$-63\pm\q9$   &$-226\pm\p13$   &&$235$ &$204$  &&0.44 &$>2.1\times10^6$ \\
SiO--6\ .....     &$\p+52\pm5$  &$+201\pm\p38$ &$+60\pm\p62$    &&$216$ &$133$  &&1.82 &$>3.7\times10^6$ \\
SiO--11\ ....     &$\p+72\pm5$  &$+137\pm\p82$ &$+146\pm\p35$   &&$213$ &$99$   &&1.57 &$>1.8\times10^6$ \\
SiO--12\ ....     &$\p+82\pm5$  &$+290\pm136$  &$-148\pm206$    &&$336$ &$74$   &&1.81 &$>1.2\times10^6$ \\
\tablecomments{
$v_x$ and $v_y$ are proper motions speeds toward the East and North, 
respectively.
$V_{total} = \sqrt{v_{LSR}^2 + v_x^2 + v_y^2}$ is the total speed of
the stars relative to \sgra.  
$V_{min}$ is the minimum total speed, calculated by subtracting $2\sigma$
for each velocity component, before calculating the speed.
Proper motions, projected distances, total velocities and enclosed mass
limits assume a distance to the Galactic Center of 8.0~kpc.
}
\enddata
\end{deluxetable}

\section{Infrared Observations}

Infrared images at K$_s$-band (2.18\micron\ wavelength) were obtained on 
2002 April 02 and May 02 on the UT4 (Yepun)
8.2-m telescope of the European Southern Observatories' VLT in Chile.  
The CONICA/NAOS adaptive optics assisted imager yielded
stellar images with up to 50\% of the
flux density contained within a full-width at half-maximum
diffraction-limited core of $\approx60$~mas.
This large-format camera produced images with an instantaneous
field of view of about a $28 \times 28$\arcs\ with 27~mas pixel size.
Even though weather conditions were not optimum 
and integration times were limited, we  
achieved a limiting magnitude near K$_s=18$~mag (\fivesig).

The April 02 data came from
a very short integration and the bright red-giant stars used for
the radio--infrared alignment were not saturated on the shift-and-add 
image.  However, this image was not well suited to detect the
weak stars within a few-tenths of an arcsec of \sgra.   
The May 02 image is shown in Fig.~4. This image used
longer integration times and went much
deeper than the April 02 image, and the weak stars near
\sgra, such as the fast moving stars S1 and S2, are 
easily detected (see Fig.~5).   The bright red-giants, with SiO 
maser emission, were somewhat saturated on this image.  However,
this only slightly degraded the positions determined for these stars
and did not markedly affect the radio--infrared alignment.

\begin{figure}
\epsscale{1.0}
\plotone{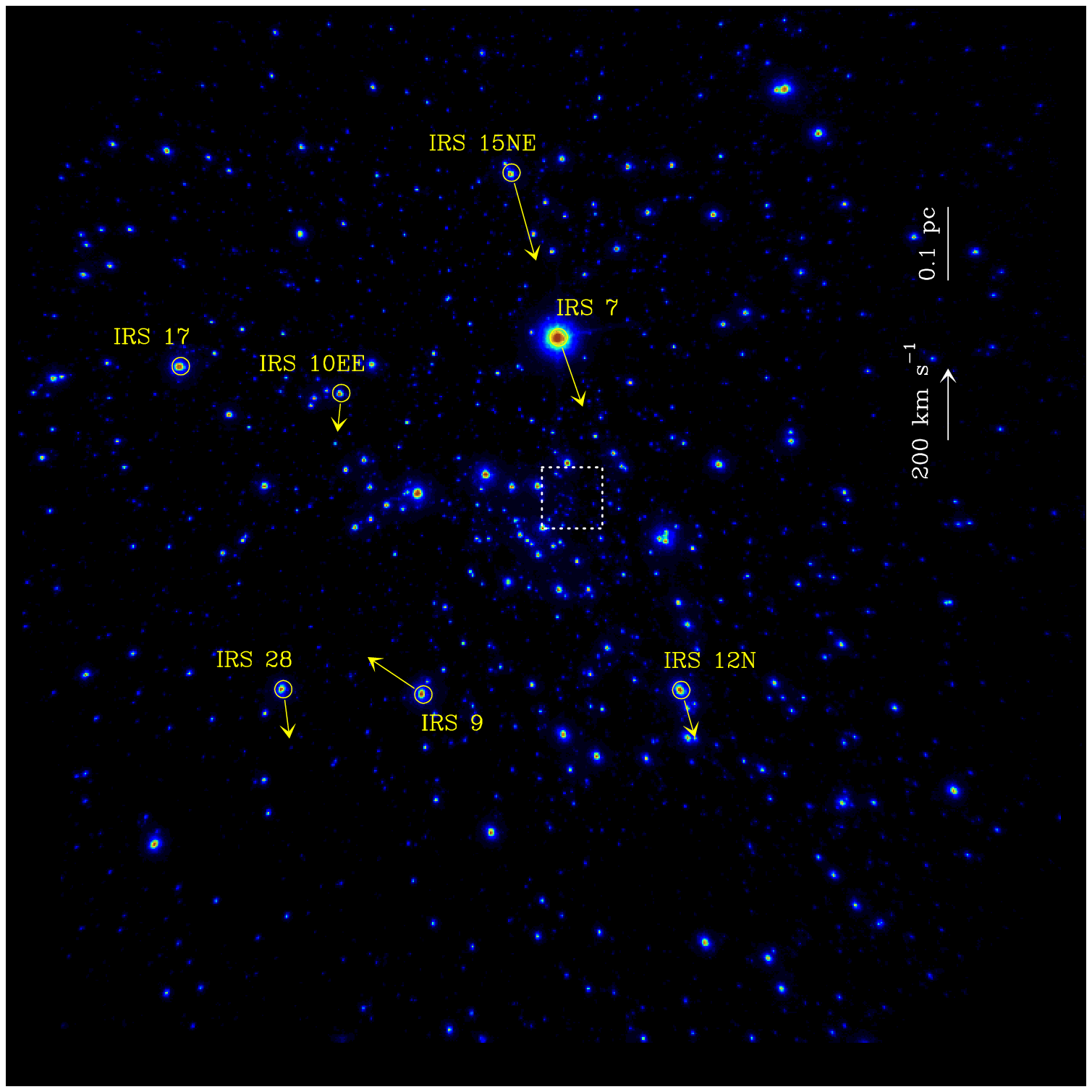}
\caption{Infrared (K$_s$-band) false-color image of the inner $\approx40$\arcs\ of
the Galactic Center region on 2002 May 02. Yellow circles locate the stars used for
the radio--infrared alignment, and arrows indicate the proper motions
of these stars {\it relative} to \sgra\ measured from SiO masers.
The white bar and arrow at the right side of the figure indicate
the linear scale of the image and the proper motion speed, respectively.
Dotted lines show the inner 2\arcs\ centered on \sgra, that is
displayed in Fig.~5.
            \label{fig4}
        }
\end{figure}

All seven red giant stars with SiO maser emission, listed
in Table~3, located within the CONICA images were
easily detected on both the April 02 and May 02 images.  
These stars are identified in Fig.~4.
In order to determine the positions of the SiO maser stars
on the infrared images, we fit an elliptical Gaussian brightness 
distribution to the diffraction-limited cores of the 
shift-and-add stellar images.  For the April 02 and May 02 images, 
$\approx10$\% and $\approx50$\%, respectively, of the flux density 
of the star was contained in these cores.  Residuals to the
fits indicated that the cores of the stellar images
were somewhat more peaked than a Gaussian function, but 
they showed no asymmetries.
Thus, the Gaussian fits should not give biased results for the
stellar positions.

Formal uncertainties in the fitted positions of these bright stars
were less than 0.5~mas.  However, given the non-Gaussian brightness
distributions of the saturated stellar images and the possibility
of anisoplanatic effects across the image, one would expect
that the true position uncertainties would exceed the formal values.
An upper limit on the combined effects of non-Gaussian distributions,
anisoplanatic effects, and other distortions can be obtained by
comparing the infrared to the more precise radio positions.  
As discussed in \S4.1, the rms deviation of the May 02 infrared 
positions, after alignment with the radio positions, was about 5~mas.
Since the alignment procedure involved 14 coordinate positions 
(for seven stars) and five free parameters, the intrinsic position
uncertainty is likely to be 6~mas.
Thus, we conclude that distortions of the infrared image 
typically are $\lax6$~mas, and that this value serves as a realistic 
estimate of the infrared position uncertainties.

\section{Radio \& Infrared Frame Alignments}

We have measured radio positions and proper motions, 
relative to \sgra, for seven bright stars within
15\arcs\ of \sgra.  The position and proper motion accuracies
currently are $\approx1$~mas and $\approx1$~\masy, respectively.
This allows us to align the radio and infrared frames, both
in position and in proper motion.
Thus, not only can the location of \sgra\ can be accurately 
determined on infrared images, but also stellar proper motions
from infrared data can be referenced directly to \sgra, without 
assumptions of isotropy or homogeneity of the stellar motions.

\subsection{Position Alignment for 2002 Images}

The infrared position fits (in pixel units) were compared to the
radio positions (J2000 east and north offsets from \sgra), corrected 
for proper motions to the epoch of the infrared observations.  
Five parameters were used to transform the infrared positions to 
best match the radio positions: two infrared image scales (in the 
east and north directions), an image rotation, and a final east and 
north translation.
These parameters were adjusted so as to minimize the difference 
between the radio and transformed infrared positions in a least-squares
sense.  A uniform joint-uncertainty of 6~mas was assigned to the
difference between the infrared and radio positions in each coordinate;
this resulted in a post-fit $\chi^2$ per degree of freedom near unity.

The least-squares fits for both infrared images produced excellent
results.   Infrared image scales were measured with formal 
uncertainties of 0.05\% and the 
east-west and north-south scales differed by less than 0.2\%.
Infrared image rotations were measured with formal uncertainties
of 0.015 degrees.  The translation parameters, which ultimately
align the infrared image with the radio positions, had formal 
uncertainties of 6 and 8~mas for the April 02 and May 02 images,
respectively.  These translation uncertainties are a direct 
indication of the {\it formal} accuracy of the radio--infrared alignment.

\begin{deluxetable}{lrrrrrrrrr}
\tabletypesize{\scriptsize}
\tablecaption   {2002 May 02 Radio--Infrared Position Alignment}
\tablehead{ \colhead{Star} &\colhead{$v_{LSR}$} 
           &\colhead{$\Delta\Theta_x^{Radio}$} &\colhead{$\Delta\Theta_y^{Radio}$} 
           &&\colhead{$\Delta\Theta_x^{IR}$}    &\colhead{$\Delta\Theta_y^{IR}$} 
           &&\colhead{$\Delta\Theta_x^{Dif}$}    &\colhead{$\Delta\Theta_y^{Dif}$} 
     \\
            \colhead{}     &\colhead{(\kms)}    
           &\colhead{(arcsec)}         &\colhead{(arcsec)}      
           &&\colhead{(arcsec)}         &\colhead{(arcsec)}      
           &&\colhead{(arcsec)}         &\colhead{(arcsec)}      
          }
\tablecolumns{10}
\startdata 
\\
IRS~9 \ .....     &$-340$  &$5.6642$  &$-6.3418$  &&$5.6639$  &$-6.3397$  &&$0.0003$  &$-0.0021$ \\
IRS~7 \ .....     &$-120$  &$0.0320$  &$5.5335$   &&$0.0290$  &$5.5449$   &&$0.0030$  &$-0.0114$ \\
IRS~12N\ ....      &$\p-63$ &$-3.2568$ &$-6.8981$  &&$-3.2571$ &$-6.8970$  &&$0.0003$  &$-0.0011$ \\
IRS~28\ .....     &$\p-55$ &$10.4688$ &$-5.8000$  &&10.4767$$ &$-5.8055$  &&$-0.0079$ &$0.0055$  \\
IRS~10EE\ ...     &$\p-27$ &$7.6847$  &$4.2069$   &&$7.6872$  &$4.2063$   &&$-0.0025$ &$0.0006$  \\
IRS~15NE\ ...     &$\p-12$ &$1.2204$  &$11.2963$  &&$1.2203$  &$11.2897$  &&$0.0001$  &$0.0066$  \\
IRS~17\ ...      &$\p+75$ &$13.1501$ &$5.5651$   &&$13.1435$  &$5.5633$  &&$0.0066$  &$0.0018$  \\  
\\
\tablecomments{Radio positions are corrected for proper motions to 
2002 May 02.  
$\Delta\Theta_x$ and $\Delta\Theta_y$ are angular offsets from \sgra\
toward the east and north, respectively, in the J2000 system.
Differenced positions (radio minus infrared) are
indicated with the superscript ``Dif''.}
\enddata
\end{deluxetable}

Table~5 summarizes the radio and transformed infrared positions
for the May 02 image.
The differences between the radio and infrared positions of the seven
stars listed in Table~5 have a root-sum-square deviation of only 5~mas.  
The largest difference is 11~mas for the notherly position offset of IRS~7.
As discussed in \S2.2, this star is the lone supergiant in our sample,
and it is likely to have an SiO maser shell size of $\approx10$~mas.  
This alone could cause the position of IRS~7, which was measured 
from the peak brightness of a single SiO emission feature, 
to be in error by $\approx5$~mas.  Also, IRS~7 is the brightest
infrared source in the field and thus was heavily saturated on
the deep image needed to detect weak emission near \sgra.
Thus, it is not surprising that the biggest difference between the 
radio and infrared positions should be for IRS~7.

The position of \sgra\ on the 2002 May 02 image is indicated on
Fig.~5, which is an expanded section of Fig.~4 showing the inner 
2\arcs\ of the Galactic Center.  Based primarily on the small 
differences between the radio and fitted infrared positions, 
we estimate that the \sgra\ infrared position has a \onesig\ 
uncertainty of $\approx10$~mas.  Note that the well studied star S2 
moved within about 16~mas of the position of \sgra\ on this
date \citep{S02}.  Thus, the emission detected near the \sgra\ position
is not that of \sgra\ itself (see \S5\ for further discussion).

Our new position for \sgra, determined from the radio/infrared alignment,
can be compared to the position of the central mass responsible for
the accelerations of stars.  Stellar accelerations measured by 
\citet{G00} and \citet{E02}, place \sgra\ about $50\pm30$~mas east and
$20\pm50$~mas south of the position determined by \citet{MREG97}.
As discussed in \S4.2, our improved infrared position for \sgra\
resulted in a shift of approximately 50~mas east and 0~mas south,
placing it well within the uncertainty of the position estimate
for the gravitational mass.  Finally, the stellar orbit determined for 
star S2 by \citet{S02} also locates the gravitational mass to within
about 10~mas of the radio/infrared position.

\begin{figure}
\epsscale{1.0}
\plotone{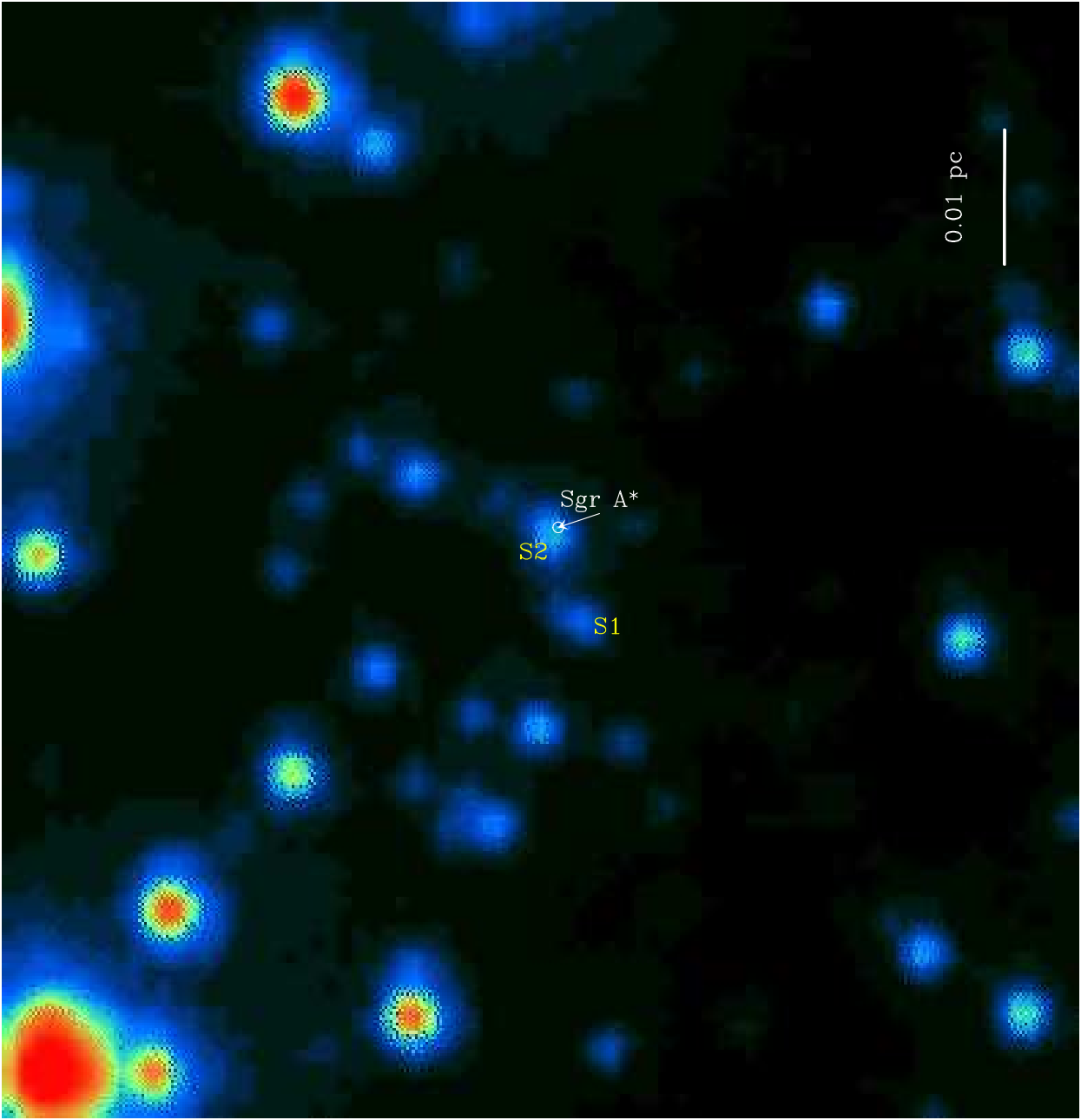}
\caption{2002 May 02 infrared image of the inner 2\arcs\ of
the Galactic Center region, corresponding to the dotted square in
Fig.~4.  The position of \sgra\ is indicated 
by the center of the small circle, whose radius of
10~mas corresponds to a \onesig\ uncertainty in the \sgra\ position.
The infrared emission close to the \sgra\ position comes from the
fast-moving star S2 projected only 16~mas from \sgra; this
star obits \sgra and was near pericenter at this time \citep{S02}.
The linear scale is indicated with the vertical line corresponding
to 0.01 pc for a distance to \sgra\ of 8~kpc.  
            \label{fig5}
        }
\end{figure}

\subsection   {Position Alignment for 1995 Images}

The radio--infrared alignment procedure of Paper~I relied primarily 
on two maser sources, IRS~7 and IRS~10EE, because of 
the small field of view of the SHARP camera ($\approx13\times13\arcs$ 
in comparison to $\approx28\times28\arcs$ for CONICA).  (A third maser
source, IRS~15NE, appeared in a mosaiced image and was only
used as a control to help evaluate accuracy.)
Owing to this limitation, only four parameters (a single pixel scale, 
a plate rotation, and a translation)
could be solved for and second order distortions could not be 
easily evaluated.  
Among the last ten years of SHARP imaging data, a small number of 
(un-mosaiced) images 
can be found that cover a section 10\arcs\ north and 5\arcs\ east of
the IRS~16 cluster. These images contain three maser sources, IRS~7, 
IRS~10EE and IRS~15NE, and thus allow a check on the influence of 
second-order image distortions. 
We found deviations of the pixel scale and rotation angle across 
the SHARP field of view can account for positioning errors which 
sometimes exceed 30~mas.  The 1995 July data used in Paper~I had
larger than average distortions. 

In order to account for second-order distortions in the SHARP images,
the following procedure was used:
\begin{itemize}
\item[] 1) A reference image is chosen and the ``reference'' positions of all
      stars are measured.  We chose a 2002 CONICA image since it has
      very small distortions and a wide field of view.           
\item[] 2) For any given epoch, preliminary positions are measured
      for all stars on the (distorted) image. 
\item[] 3) A model position for each star, $(x_m,y_m)$, is generated from the
      reference positions, $(x_r,y_r)$, and parameterized image 
      transformation equations, which account for plate scale, rotation, 
      distortion (to second order) and translation \citep{EG97}.
      The transformation equations are as follows:
      $$x_m = a_0 + a_1 x_r + a_2 y_r + a_3 x_r^2 + a_4 x_r y_r + a_5 y_r^2$$
      $$y_m = b_0 + b_1 x_r + b_2 y_r + b_3 x_r^2 + b_4 x_r y_r + b_5 y_r^2.$$
      The parameters, $a_i$ and $b_i$, are adjusted to minimize the differences
      between the preliminary and model positions in a least-squares sense.
      Since the number of parameters for a given image
      is much smaller than the number of stars, this can work very well.
\end{itemize}

After correcting SHARP data for second-order distortions,
a new alignment of the radio and the infrared reference frames results in 
a position for \sgra\ on the 1995 July SHARP image that is about 
50~mas east of the location determined in Paper~I.  This position
still lies within the \twosig\ error circle of 60~mas estimated in Paper~I.
Only a small fraction of this change is the result of improved radio 
positions.  Most of this change comes from correcting for second-order 
distortions across the SHARP image.  We estimate that the \onesig\ 
uncertainty in the location of \sgra\ is 15~mas for this image.  
Note that the position of source S3 
now appears $\approx$\twosig\ offset from that of \sgra.

\begin{figure}
\epsscale{1.0}
\plotone{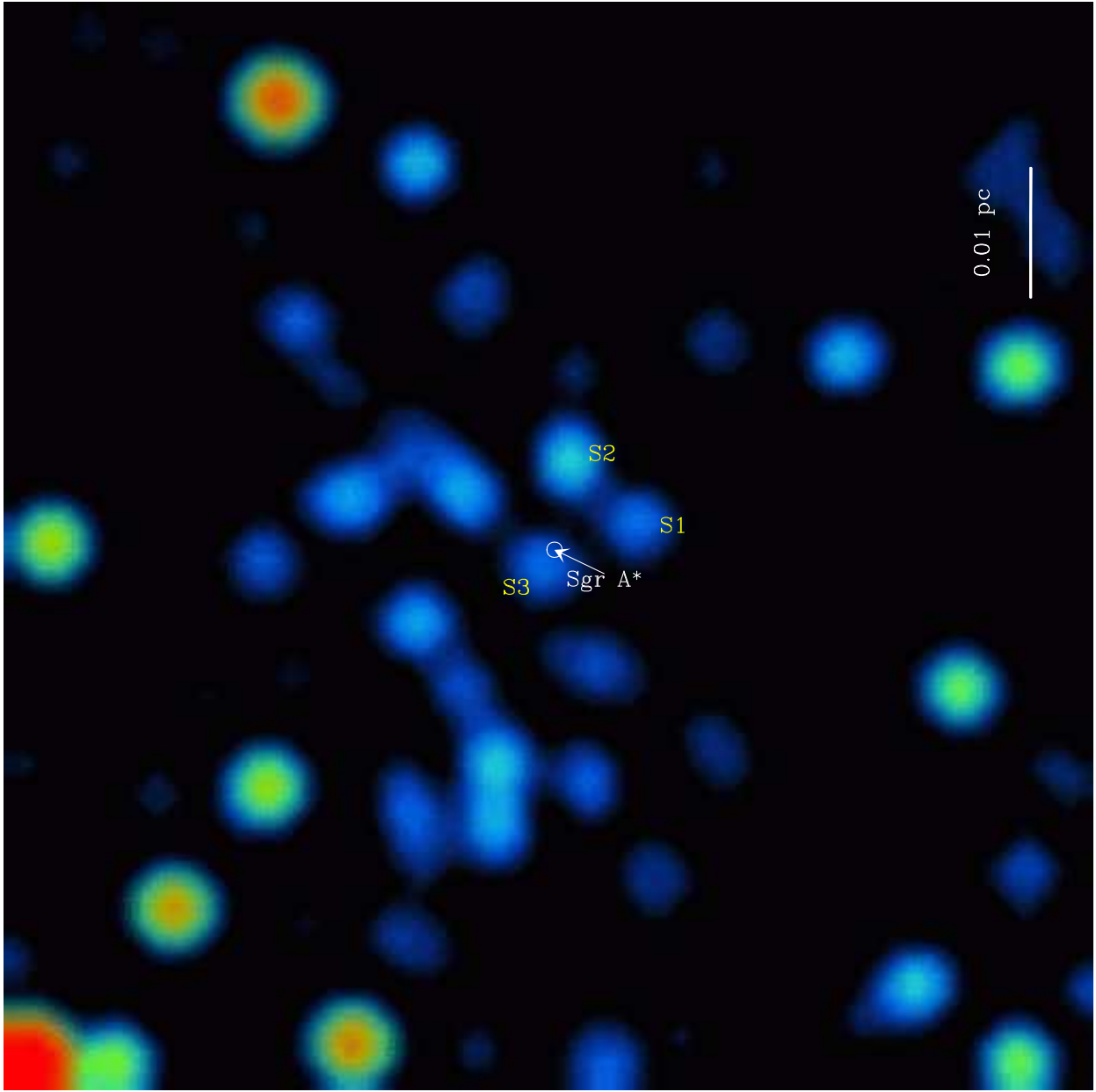}
\caption{1995 July infrared image of the inner 2\arcs\ of
the Galactic Center region, corresponding approximately to the
region shown in Fig.~5 for 2002 May.  The new position of \sgra,
corrected for second-order image distortions in the SHARP system, 
is indicated by the center of the circle, whose radius of
15~mas corresponds to the estimated \onesig\ uncertainty in the \sgra\ position.
The infrared emission close to the \sgra\ position comes from the
star S3, projected only $\approx30$~mas from the \sgra\ position.
The linear scale is indicated with the vertical line corresponding
to 0.01 pc for a distance to \sgra\ of 8~kpc.  
            \label{fig6}
        }
\end{figure}

\citet{G00} and \citet{E02} used
acceleration measurements of stars close to \sgra\ in order to locate
the massive black hole candidate.  Both groups found
an offset of $\approx50$~mas to the east of the nominal radio position of Paper~I.
With the new alignment of the radio and
IR frames, there is now good agreement between the radio--infrared
position of \sgra\ and its position inferred from stellar accelerations.

\subsection{Proper Motion Reference Frame}

The proper motions of stars in the Galactic Center cluster, measured
from infrared images, are {\it relative} motions only.  One can
add an arbitrary constant vector to all of the stellar 
proper motions without violating observational constraints.
Until now, the ``zero points'' of the motions have been determined by 
assuming isotropy and removing the average motion of the entire
sample.   Since the radio proper motions are inherently in a
reference frame tied directly to \sgra, one can use any one of 
the SiO stars, or the mean motion of a group of them,
to place the infrared proper motions in \sgra's frame.

Radio and infrared proper motions have been measured for the four  
stars located within a projected distance of 10\arcs\ on the 
sky of \sgra: IRS~9, IRS~7, IRS~12N, and IRS~10EE.  These motions
are listed in Table~6, along with the differences of the radio
and infrared motions.  
The four stars have unweighted mean differences (and standard
error of the means) of  
$+0.84\pm0.85$~\masy\ toward the east, and
$-0.25\pm0.96$~\masy\ toward the north.
One star, IRS~12N, has a significant discrepancy between the radio and 
infrared motions in the both coordinates.  Possibly the infrared
motion of this star suffers from blending of multiple stars (within the
VLT resolution).  If this source is removed, the 
unweighted mean differences become   
$+0.30\pm0.87$~\masy\ toward the east, and
$+0.36\pm1.00$~\masy\ toward the north. 
Using the results from either four or three stars, 
the infrared motions, corrected to zero 
mean motions, are consistent with being in a reference frame tied to 
\sgra, within uncertainties of $\approx1$~\masy\ or $\approx40$~\kms.
We conclude that, on average, the stars in the Galactic Center
cluster have no net motion with respect to \sgra\ at the
$\lax40$~\kms\ level per coordinate axis or 
$\lax70$~\kms\ for a 3-dimensional speed.

\begin{deluxetable}{lrrrrrrrr}
\tabletypesize{\scriptsize}
\tablecaption   {Radio--Infrared Proper Motion Alignment}
\tablehead{ \colhead{Star}  
           &\colhead{$\mu_x^{Radio}$} &\colhead{$\mu_y^{Radio}$} 
           &&\colhead{$\mu_x^{IR}$}    &\colhead{$\mu_y^{IR}$} 
           &&\colhead{$\mu_x^{Dif}$}    &\colhead{$\mu_y^{Dif}$} 
     \\
            \colhead{}      
           &\colhead{(\masy)}         &\colhead{(\masy)}      
           &&\colhead{(\masy)}         &\colhead{(\masy)}      
           &&\colhead{(\masy)}         &\colhead{(\masy)}      
          }
\tablecolumns{9}
\startdata 
\\
IRS~9 \ .....     &$+3.60\pm0.53$ &$+2.40\pm1.13$  &&$+1.99\pm0.40$ &$+0.51\pm0.24$ &&$+1.61\pm0.66$  &$+1.89\pm1.15$ \\
IRS~7 \ .....     &$-1.57\pm0.92$ &$-4.54\pm3.47$  &&$-0.75\pm0.23$ &$-3.63\pm0.48$ &&$-0.82\pm0.95$  &$-0.91\pm3.50$ \\
IRS~12N\ ....     &$-0.82\pm0.22$ &$-2.81\pm0.26$  &&$-3.26\pm0.45$ &$-0.76\pm0.81$ &&$+2.44\pm0.50$  &$-2.05\pm0.85$ \\
IRS~10EE\ ...     &$+0.20\pm0.16$ &$-2.12\pm0.20$  &&$+0.08\pm0.32$ &$-2.21\pm0.93$ &&$+0.12\pm0.36$  &$+0.09\pm0.95$ \\
\\
\tablecomments{$\mu_x$ and $\mu_y$ are proper motions relative to \sgra\
toward the east and north, respectively.
Differenced motions (radio minus infrared) are
indicated with the superscript ``Dif''.
Radio motions are in a reference frame tied to \sgra; infrared motions
are relative motions, with the average removed.
The un-weighted mean motion difference is 
$+0.84\pm0.85$~\masy\ toward the east, and
$-0.25\pm0.96$~\masy\ toward the north.
}
\enddata
\end{deluxetable}

\section{Infrared Emission of \sgra}

As mentioned in \S4.1, the well-studied star S2 moved within 16~mas of \sgra\
near pericenter passage early in 2002 \citep{S02}.  While S2 was just
beginning to move away from \sgra\ by 2002 May 02, when the
CONICA image shown in Fig.~5 was taken, it was still well within the
diffraction-limited beam of the VLT telescope.  Thus, even though
\sgra's position is known to 10~mas accuracy at this time,
one cannot use this data to determine the infrared emission 
of \sgra.   Later in 2002, or more likely in 2003, when S2 has 
moved sufficiently far away from \sgra, it should be possible
to do this. Currently the best limit for {\it steady} infrared emission from
\sgra\ is 2~mJy \citep{HG02}.

\section{Enclosed Mass versus Radius from \sgra}

Previous estimates of the enclosed mass versus projected radius from
\sgra, based on infrared stellar motions, rely on {\it relative} motions 
not tied directly to \sgra.
Since, the proper motions of the SiO masers in this paper are directly
tied to \sgra, they provide a valuable check on the enclosed mass within 
a projected radius of 0.2 to 2 pc of \sgra.

We now estimate a lower limit to the enclosed mass at the
projected radius of each star, assuming the stellar motions reflect gravitational 
orbits dominated by a central point mass. Given the 3-dimensional speed, $V_{total}$, 
and projected distance from \sgra, $r_{proj}$, for each star, one can determine a 
strict lower limit to the mass enclosed, $M_{encl}$, within the true radius, $r$, 
of that star from \sgra.  For a given mass and semi-major axis, the greatest orbital
speed occurs at pericenter and is given by the well known equation
$$V_{pc}^2 = {GM_{encl}\over a}{\bigl({1+e\over1-e}\bigr)}~~,~~\eqno(1)$$
where $G$ is the gravitational constant, $a$ is the orbital semi-major axis,
and $e$ is the orbital eccentricity.   The pericenter distance is
given by $r_{pc} = a(1-e)$ and, for any orbit,
the projected pericenter distance cannot exceed the true distance:
$r_{proj} \le r_{pc}$.  Finally, since $V_{total} \le V_{pc}$, Eq.~(1) yields 
$$V_{total}^2 \le {GM_{encl}(1+e)\over r_{proj}}~~.~~\eqno(2)$$
This inequality is maintained by setting $e=1$, and Eq.~(2) yields
a strict lower limit to the central mass:
$$M_{encl} \ge {V_{total}^2 r_{proj} \over 2G}~~.~~\eqno(3)$$
This lower limit approaches an equality only when $r_{proj} \rightarrow r$,
the star is near pericenter, and it has an eccentricity near 1.
 
We evaluate the lower limit to $M_{encl}$ using Eq.~(3) and adopting
the smallest total velocity allowed by our $2\sigma$ measurement uncertainties.   
These mass limits, given in Table~4,
are mostly consistent with the enclosed mass versus projected distance from 
\sgra\ given by \citet{GEOE97} and \citet{G98}.
For many of the stars, the lower mass limits are well below the estimated
enclosed mass curves, as expected given the conservative nature of the 
calculated limits.  However, since $r_{proj} \ge 0.9\times r$ 
about 40\% of the time (for random distributions), we would expect to 
have some significant mass limits from a sample of 9 stars.  
Indeed, IRS~9, IRS~15NE, and SiO--6 have lower limits to the
enclosed mass that are close to 
the values estimated from infrared stellar motions.  

Our most significant lower mass limit is from IRS~9: $>4.5\times10^6$~\msun\ 
at a projected radius of 0.33 pc from \sgra. 
This mass exceeds by 50\% the currently favored model of a $2.6\times10^6$~\msun\ 
black hole, plus a $0.4\times10^6$~\msun\ contribution from the central
stellar cluster at this radius \citep{GEOE97,G98}. 
As pointed out above, this is an extremely conservative limit, since it assumes
1) an orbital eccentricity near 1.0, 2) the star is currently near pericenter 
(where it spends little time), 3) $r_{proj} \approx r$, and 
4) the 3-dimensional speed is over-estimated by its $2\sigma$ uncertainties. 
Thus, either IRS~9 is {\it not} bound to \sgra, or current 
models underestimate the enclosed mass at a radius of about 0.3~pc from \sgra.

\acknowledgments
M. Reid thanks the Alexander von Humboldt Foundation for support
while at the Max-Planck-Institut f\"ur Radioastronomie where
a portion of this work was completed.  We thank A. Beasley
for providing a special AIPS task (PHREF), which was needed to analyze the
VLBA data.  We also thank A. Ghez for a comments on the paper.

\clearpage

\end{document}